# NeuroSleepNet: An Explainable Multi-Head Attention-Based Framework with Spatial and Multi-Scale Independent Temporal Context Learning for Automatic Sleep Stage Scoring

**Muhammad Sudipto Siam Dip[1], Student Member, IEEE, Mohammod Abdul Motin[1], Senior Member, IEEE, Chandan Karmakar[2], Member, IEEE, Thomas Penzel[3], Fellow, IEEE, and Marimuthu Palaniswami[4], Life Fellow, IEEE**

[1]Department of Electrical & Electronic Engineering, Rajshahi University of Engineering & Technology, Bangladesh
[2]School of Information Technology, Deakin University, Geelong, Victoria, Australia
[3]Interdisciplinary Sleep Medicine Centre, Charité University Hospital, Berlin, Germany
[4]Department of Electrical & Electronic Engineering, The University of Melbourne, Parkville, Melbourne, VIC 3010, Australia

Corresponding author: Chandan Karmakar (e-mail: karmakar@deakin.edu.au).

**ABSTRACT** Automatic sleep stage scoring is crucial for monitoring sleep quality and sleep-related disorders. However, most existing electroencephalogram (EEG) based deep learning frameworks rely on current and previous input sequences (varying from 4–100 epochs), which increases computational complexity and creates dependency on transition rules. In this study, we propose NeuroSleepNet, a deep-learning framework that classifies sleep stages using only the current input sequence. NeuroSleepNet employs a two-stage representation learning that includes spatial and multi-scale independent temporal context learning. Multi-head self-attention transformer encoders were employed to transform extracted representations into contextual embeddings to improve the distinctiveness of learned features. Additionally, to handle the class imbalances in different sleep stages, a logarithmic scale-based weighting technique was incorporated into the loss function of NeuroSleepNet. Furthermore, an explainability module was integrated into NeuroSleepNet to visualize and interpret class-specific decision patterns. NeuroSleepNet demonstrated state-of-the-art performance, achieving an accuracy of 86.1%, a macro-F1 score of 80.8%, and a Cohen's kappa of 0.805 on the Sleep-EDF expanded dataset. It achieved 82.0%, 76.3%, and 0.753 on MESA, 80.5%, 76.8%, and 0.738 on Physio2018, and 86.7%, 80.9%, and 0.804 on the SHHS database, respectively, by utilizing only the current sequence. As a result, there is no need for previous long sequences or sequential dependencies, making it computationally efficient and well-suited for long-term automatic sleep monitoring.
**Source code:** https://github.com/SiamDip/NeuroSleepNet

**INDEX TERMS** Electroencephalogram signal, multi-scale temporal context learning, sleep stage classification, spatial learning, transformer encoder, weighted loss function

## I. INTRODUCTION

Sleep plays a critical role in maintaining cognitive function and overall mental health. Inadequate sleep has been linked to impairments in memory, concentration, and emotional regulation [1]. Understanding and correctly identifying sleep stages is crucial for monitoring sleep quality and diagnosing sleep-related disorders [2]. Clinical sleep studies rely on the Polysomnography signals (PSG), which consist of multiple physiological signals such as the Electroencephalogram (EEG), Electromyogram (EMG), Electrocardiogram (ECG), and Electrooculogram (EOG). Sleep stage scoring is classified into wakefulness (W), rapid eye movement (REM), and three distinct non-REM (NREM) periods, such as N1, N2, and N3. These stages are conventionally scored on 30-second segments as defined by the R&K and AASM guidelines [3, 4]. The occurrence of these stages varies overnight [4]. EEG-focused approaches have previously achieved performance comparable to gold-standard multi-modal PSG-based sleep staging, while offering greater practicality for wearable and home-based sleep monitoring applications. The early studies of automatic sleep scoring involved the incorporation of hand-crafted features from various domains and the use of traditional classifiers [5-7]. To address the need for automated learning of complex patterns from raw data, various end-to-end deep learning frameworks have been developed for automated sleep-stage classification from EEG signals. Convolutional neural networks (CNNs) have demonstrated their effectiveness as a powerful tool for learning representations, which can be integrated with fully connected neural networks [8]. Besides CNN-based architectures, recurrent neural networks (RNNs) have been utilized due to their effectiveness with sequential data [9]. Recently, hybrid models of CNNs with RNNs [10-15] have become very popular due to their ability to leverage the strengths of both CNNs in capturing spatial features and RNNs in handling temporal dependencies, making them highly effective. Other methods, such as transformer [16], CNN with transformer [17-20], image-representation-based methodologies [21], joint representation [22], and transfer learning [23] are also explored in this field.

Among these studies, multi-scale convolutional approaches are widely adopted due to their ability to capture both temporal and spectral features through varied receptive field Sizes. Previous studies have shown that small filters are more effective in extracting temporal changes, while large filters help in extracting the frequency contents of EEG signals [24]. An early well-known benchmark on utilizing this principle in sleep-stage scoring was introduced by Supratak et al. in [10] and further investigated in [19-21, 25-27]. However, despite their strong performance, these models depend heavily on long temporal sequences and inter-epoch dependencies.

In contrast, while traditional methods using hand-crafted features can be optimized for better performance with shorter sequence lengths [5], deep learning-based sleep staging models frequently rely on long input sequences to capture transitional patterns consistent with established sleep transition rules. Although extending sequence length enhances temporal modeling, it also substantially increases memory requirement and computational cost [28]. Models that incorporate both past and future epochs introduce latency, limiting their suitability in real-time settings. In contrast, compact and low-latency sleep-stage scoring models are better suited for deployment on wearables or mobile devices [29]. Moreover, real-world sleep is often irregular since many individuals experience fragmented sleep patterns, limited deep sleep (N3), or an absence of REM. Conditions like sleep apnea further disrupt typical stage transitions, making long-term patterns unreliable. In such cases, models that rely heavily on extended context or a full sleep cycle may struggle [30]. Previous studies [21, 25, 27, 31] attempted to reduce the sequence length used for sleep stage classification but observed decreased performance, particularly for stages of N1 and REM. We argue that reducing sequence dependency and producing models that depend solely on the current epoch can increase efficiency and make deep learning-based sleep staging more suitable for clinical real-world applications. Additionally, explainability in these deep learning-based sleep staging is less explored, which involves understanding how the model distinguishes between sleep stages and the reasoning behind its classification decisions. In sleep staging, explainability often relates to how the model assigns importance to specific signal regions, particularly in cases where stage boundaries are ambiguous, which is equally important for transparency [32]. Beyond these challenges, reducing computational parameters in deep learning models is also a notable consideration as it enhances efficiency, lowers memory requirements, and makes sleep staging systems more practical for faster as well as suitable for resource-constrained settings.

To overcome the above-mentioned constraints, we proposed NeuroSleepNet, a multi-head attention-based deep learning model that progressively captures spatial and temporal information and predicts sleep stages using only intra-epoch temporal contexts of a 30 seconds EEG segment. The concept of contextual embedding was employed to capture meaningful relationships within intra-epoch features. Four different datasets were used to investigate the performance of our model. The model is designed with efficient representation learning, which makes it a more compact overall architecture. Above all, we incorporated explainability techniques to explain the model's decision-making. The major contributions of our study are outlined below:

- A novel deep learning architecture, NeuroSleepNet, is proposed to monitor automated sleep staging from EEG, integrating deep learning-based spatial filtering, multi-scale independent temporal representation learning, and a multi-head self-attention transformer encoder.
- Most previous state-of-the-art studies rely not only on the current sequence but also on a large number of previous sequences. In contrast, our primary objective was to classify sleep stages using only the current epoch without referencing previous epochs. This approach enhances model efficiency and eliminates dependency on prior epochs or transition rules for automated sleep stage classification.
- A weighted logarithmic-scaled approach is integrated into the loss function to address class imbalance across different sleep stages. Additionally, an explainability mechanism was incorporated into the proposed model to enhance interpretability and provide insights into the decision-making process for sleep stage classification.

The proposed NeuroSleepNet model was validated using four publicly available datasets, and it demonstrated state-of-the-art performance.

## II. NEUROSLEEPNET

NeuroSleepNet was developed in two stages to keep it generalizable and reduce the risk of overfitting. In the initial design stage, we designed the core architecture based on fixed design choices, such as the type and formation of modules, transformer token arrangement, kernel scaling strategies, and fixed filter counts. These decisions were made before testing on any dataset to keep the model mostly independent of specific data characteristics. In the subsequent tuning stage, only two components were fine-tuned on our primary dataset, e.g., the smallest multi-scale independent temporal context learning (*MITCL*) kernel size and transformer settings. Since only a small number of parameters need tuning, the model remains simple and transferable to other datasets. The rest of this section focuses on the initial architectural design, while the choices of tunable parameters are discussed in III.

**Input Formulation:** The dataset is constructed with 30-second EEG segments sampled at 100Hz. Here, $X = \{x_1, x_2, \dots x_n\}$ represents the *n* number of EEG segments in which $x_i = \mathbb{R}^{C \times T}$. The *C* represents the number of channels, and *T* is the total samples per segment. Corresponding to $x_i$, an encoded label is defined as $y_i \in \{0,1\}^Y$ where *Y*=5, indicating the set of five stages of sleep, notably $\{W, N1, N2, N3, R\}$. In our standard one-to-one approach, NeuroSleepNet takes only the current segment $x_i$ to predict its corresponding label $y_i$.

### A. CONVOLUTIONAL SPATIAL FILTERING

The Convolutional spatial filtering (*CSF*) concept in this paper was inspired by [33], which shares a similar motivation to previous studies mentioned in [34]. Here, the aim is to create a convolutional block that can process all channels simultaneously at each time point. This approach enables the model to capture cross-channel features, thereby enhancing subsequent analysis. The convolutional block performs a linear operation on the multi-channel input, leading to the generation of virtual channels. Spatial filtering was performed using a 2D convolutional block with a number of filters equal to the input channels and a kernel size of *(C, 1)*.

This operation produces two vectors corresponding to two channels, preserving the input length *T* of the raw signal, commonly referred to as virtual channels.

This was applied using 2D convolution, which takes a 2D input of *(1, C, T)* reshaped from (*C, T*), where *C* is the number of channels indicating input height $h_i$ and *T* is samples per segment indicating the width of the input $w_i$. Since the filter size is $(\kappa_h, \kappa_w)$ = *(C, 1)*, we can indicate that the stride height $h_s$ is equal to 1 and the stride width $w_s$ is equal to 1. Thus, the output height is equal to 1, and the output width is equal to *T*.

Therefore, combining every channels, the output is *(*1, *T)* and since the number of filters is the same as the number of channels *C*, we get the final output of the spatial block as *(*|B|, *C,* 1, *T)*, where |B| is the input batch size. The output is permuted back to (|B|, 1, *C*, *T*) to get back to its original input shape to maintain consistency with the channel dimension.

### B. MULTI-SCALE INDEPENDENT TEMPORAL CONTEXT LEARNING

NeuroSleepNet employs convolutional blocks of various scales concurrently to extract the temporal context from the virtual input channels, defined as multi-scale independent temporal context learning (*MITCL*). Unlike traditional temporal context learning implemented with 1D CNN, which aggregates information across multiple EEG channels through element-wise multiplication and summation, each convolutional block in *MITCL* applies convolutional operations independently to each channel. This design preserves the distinct temporal dynamics of each virtual channel while extracting the temporal information. In *MITCL*, we denote the number of convolutional blocks in parallel as *P*. Each block includes several convolutional filters, a rectified linear unit (*ReLU*) for activation, and a max-pooling layer to down-sample the convolution block's outputs. The complete operation of *MITCL* can be shown as

$$CNN_{MITCL} = \{ CNN_{\theta_1}(.), \dots CNN_{\theta_P}(.)\} \quad (1)$$

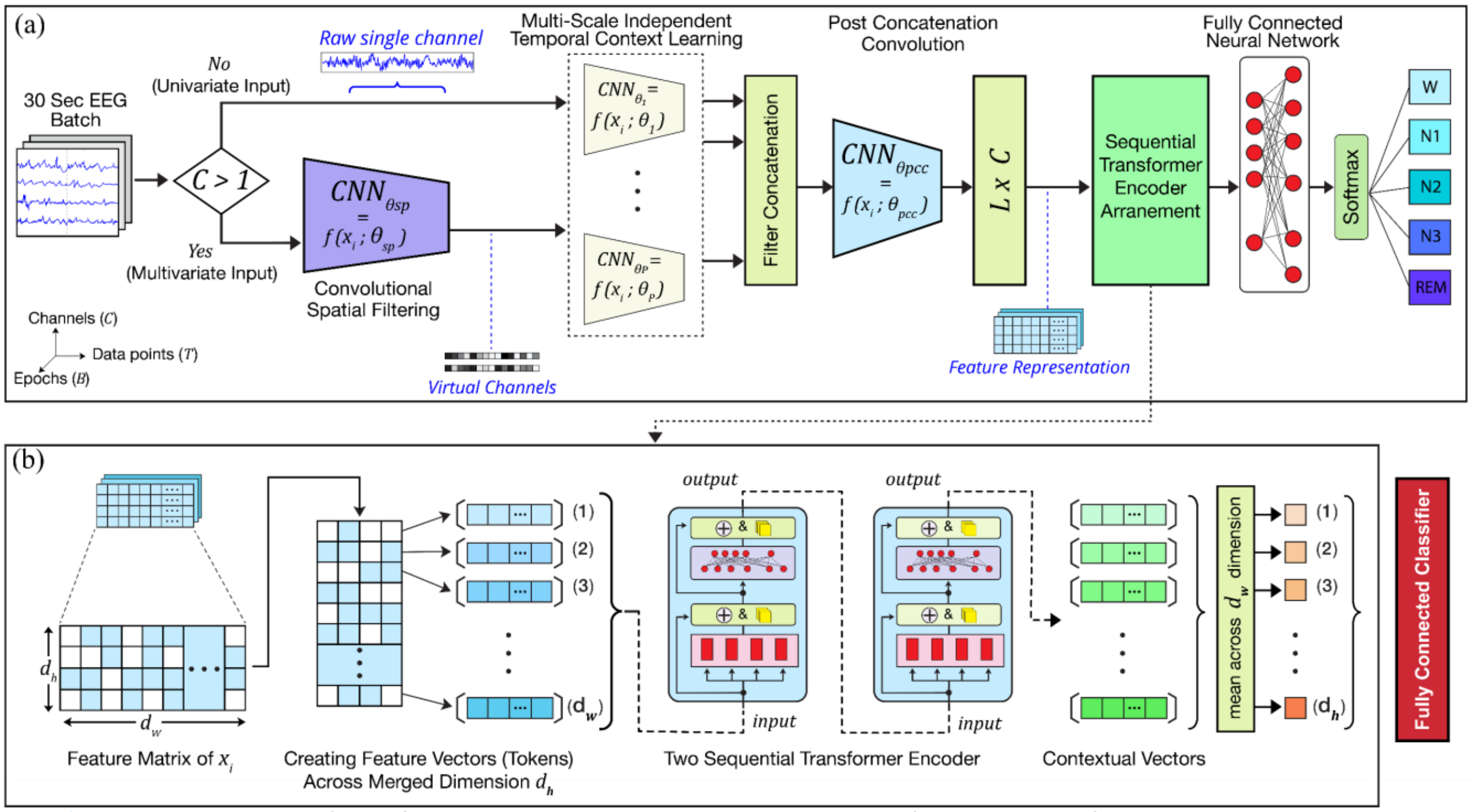

**FIGURE 1.** a) The overall architecture of NeuroSleepnet: the initial representation learning to extract features consists of spatial and multi-scale independent temporal context learning, which is then concatenated across the filter dimension, followed by a post concatenation convolution block. Lastly, the filter and additional channel dimension are then merged and fed into the transformer encoder arrangement, and lastly into the fully connected neural network classifier. b) The transformer encoder arrangement is fully demonstrated. From a feature matrix, token vectors are created to refine feature representations and transform them into contextual features using two sequential transformer encoders and passing them into the classifier. Here, *xi* represents the input segment, and *θ* represents the trainable parameters.

TABLE I
MODEL SPECIFICATION OF NEUROSLEEPNET'S REPRESENTATION LEARNING. THE FEATURE EXTRACTOR IS DIVIDED INTO THREE STAGES: 1) SPATIAL LAYER, 2) MITCL, AND 3) PCC

| Layer name | Layer type | Filters(L) | Filter size | Stride | Output shape | Activations | Paddings |
|---|---|---|---|---|---|---|---|
| SPATIAL LAYER | Input | -- | -- | -- | (C, T) | -- | -- |
| | Reshape | -- | -- | -- | (1, C, T) | -- | -- |
| | Conv 2D | C | (C, 1) | (1, 1) | (C, 1, T) | Linear | -- |
| | Permute | -- | -- | -- | (1, C, T) | -- | -- |
| MITCL | Conv2D<br>Maxpooling 2D (i=1) | 8 | (1, $K_1$*i)<br>(1, 12) | (1, 1) | (8, C, T)<br>(8, C, T//12) | ReLU | $K_1$/2 |
| | . . . | . . . | . . . | . . . | . . . | . . . | . . . |
| | Conv2D<br>Maxpooling 2D (i=P) | 8 | (1, $K_P$ = $K_1$*P )<br>(1, 12) | (1, 1) | (8, C, T) | ReLU | $K_P$/2 |
| | Filter Concatenation | -- | -- | -- | (8*P, C, T//12) | -- | -- |
| PCC | Conv 2D<br>Maxpooling 2D | 8 | (1, $K_1$*i)<br>(1, 12) | (1, 1) | (C, 1, T) | ReLU | $K_1$/2 |
| | Dimension Fusion | -- | -- | -- | (8*C, T//(12*12)) | -- | -- |
| | Permute | -- | -- | -- | (T//(12*12), 8*C) | -- | -- |

Here, $CNN_{\theta_j}(.)$ indicates each convolutional block for independent temporal context learning, and $\theta_j$ represents the parameter of the $j^{th}$ CNN, indicating its distinct filter size. From the feature matrix created as virtual channels *X'*, the *MITCL* takes $x_i'$ and outputs a series of feature arrays denoted as

$$\mathrm{C}_{t,i} = \{\mathrm{C}_t^{\theta 1}, \mathrm{C}_t^{\theta 2} \dots \mathrm{C}_t^{\theta P} \} \quad (2)$$

The 2D convolutional block is created into an independent temporal convolution layer by setting the kernel size $(\kappa_h, \kappa_w)$ as (1, *K*). The first dimension of the kernel for each convolutional block is kept 1 to apply the same filter across the spatial dimension and avoid the element-wise multiplication & summation. By applying $L$ number of filters, each convolutional block outputs $L$ number of feature matrices shaped $(h_{mtcl}, w_{mtcl})$. Here, the $h_{mtcl}$ is equal to C as padding *p* is 0, and $w_{mtcl}$ is kept the same length as *T* by adjusting the padding size as:

$$p = \frac{\kappa_w}{2} = \frac{K}{2} \quad (3)$$

The kernel width *K* is taken differently for different blocks in the *MITCL* operation, which is progressively larger convolutional kernels (e.g., each twice the size of the previous). Therefore, if the smallest kernel size or the kernel size of $CNN_{\theta_1}(.)$ is $k_1$, then the set of kernel values for $CNN_{MITCL}$ follows:

$$K_{MITCL} = \{ k_1, k_1 * 2 \dots, k_1 * P\} \quad (4)$$

Each of the convolutional blocks in *MITCL* downsamples with a max-pooling layer having a pool of $r$ temporal samples. Thus, every feature array results in $\mathrm{C}_t^{\theta j} \in \mathbb{R}^{L \times C \times [T/r]}$. Each parameter is depicted in Table I. Following the *MITCL*, we performed filter concatenation to each output of $\mathrm{C}_t$ across the dimension of filter that can be denoted as

$$\mathrm{C}_{t(cat),i} = [\, \mathrm{C}_t^{\theta 1} || \, \mathrm{C}_t^{\theta 2} || \dots || \, \mathrm{C}_t^{\theta P} \,]_{dim=1} \quad (5)$$

The number of filters after filter concatenation is $L' = P.L$. Thus, the concatenated array is $\mathrm{C}_{t(cat),i} \in \mathbb{R}^{(L.P) \times C \times [T/r]}$. This concatenated array is passed to a post-concatenation convolution (*PCC*) block with *ReLU* activation and a max-pooling layer and to simplify the combined representations coming out of the *MITCL*. The parameters of *PCC* were kept similar to the first convolutional block of *MITCL,* denoted as $CNN_{\theta_1}(.)$, which had the smallest filter scale among all. The output of this *PCC* is $\mathrm{C}_{PCC,i} \in \mathbb{R}^{L \times C \times [T/(r.r)]}$. The benefit of this operation is further discussed in section IV-D. The filter dimension and channel dimension are merged into a single dimension by multiplying them, and the result is denoted as $\mathrm{C}_{out,i} \in \mathbb{R}^{(L.C) \times [T/(r.r)]}$, where $\mathrm{C}_{out,i}$ represents the final feature matrix of the $i^{th}$ segment from the mini-batch in our representation learning. The output from here is permuted again and prepared for the transformer encoder layer. The overall formation of NeuroSleepNet's modules is simplified in Fig. 1(a).

matrix of the $i^{th}$ segment from the mini-batch in our representation learning. The output from here is permuted again and prepared for the transformer encoder layer. The overall formation of NeuroSleepNet's modules is simplified in Fig. 1(a).

### C. SELF-ATTENTION BASED MULTI-HEAD ENCODER

To enhance the representation of extracted features and capture more complex dependencies, we employed a multi-head self-attention-based *(MHSA)* transformer encoder in our model [35]. The self-attention mechanism enables the model to transform static feature representations into contextual ones by dynamically weighing their relevance to surrounding sequences. Additionally, the multi-head structure allows the model to attend to different aspects of the input simultaneously, capturing diverse patterns and relationships within the data. In this study, we used $N = 2$ identical encoders sequentially. Each encoder consists of four self-attention blocks, i.e., $n_{head} = 4$. The encoder takes a feature matrix of $C_{out} \in \mathbb{R}^{d_h \times d_w}$, where $d_h = (L.C)$ represents merged dimensions of *PCC* filters, and EEG channels and $d_w = [T/(r.r)]$ represents pooled temporal sequence length, denoted for simplification for the later stages. The input of the encoder network takes a sequence, $\rho$, which contains $d_w$ number of token vectors where each vector consists of $d_h$ values. The input of an encoder network can be easily related to that of natural language processing, in which we can consider $d_w$ as the number of words embedding vectors, where each one consists of $d_h$ number of values representing $d_h$ distinct characters. Thus, $d_h$ is considered the shape of $d_{model}$ *a.k.a.* embedding dimension of the transformer encoder. Fig. 1(b) shows a simplified representation of the data arrangement for the transformer encoder and the following steps.

After processing with four self-attention blocks, an addition operation with the original input sequence and layer normalization is applied, following the original settings from [35]. The second part of the encoder network uses a default two-layered feed-forward neural network where the hidden layer consists of *2048* units with *ReLU* activation function and a dropout of 0.25. The following layer outputs a series of vectors where each vector has $d_{model}$ number of values for each input sequence. Formally, it can be written as, $z = \{z_1, z_2, \dots z_n\}$. The output is then again added with the previous inputs- the normalized vectors, and normalized again with 2nd layer normalization. The final output can be depicted as:

$$\mathcal{Y} = TransformerEncoder(C_{out,i}, \theta_e) \quad (7)$$

Where $\mathcal{Y} = \{y_1, y_2, \dots y_n\}$ and $\mathcal{Y} \in \mathbb{R}^{d_h \times d_w}$ indicates the same shape of the input of the transformer encoder. The output $\mathcal{Y}$ is then flattened by averaging across their temporal dimension $d_w$ using the following relation:

$$A_i = \frac{1}{d_w} \sum_{j=1}^{d_w} d_{h_j} \quad (8)$$

Where, $A_i \in \mathbb{R}^{1 \times d_h}$. This flattened vector representing the $i^{th}$ segment from a mini-batch is fed to a single-layered neural network classifier with $L.C$ number of hidden units, followed by a *softmax* layer to predict the sleep stage.

### D. LOG-SCALED WEIGHTED COST FUNCTION

The cross-entropy loss for a multi-class classification task is a loss function to measure the performance of a classification model. It measures the discrepancy between the actual probability distribution found by a classification model and the predicted values. In a conventional way, the cross-entropy loss is computed from a mini-batch of data containing $n$ segments; the model predicts the output logits shaped ($n$, $Y$). Each of these logits is then converted into probabilities with this softmax expression:

$$p_{i,Y} = \frac{\exp(x_{i,Y})}{\sum_{j=1}^{Y} \exp(x_{i,j})} \quad (9)$$

Where $i$ indicates the sample index and $Y$ is the number of classes. From the matrix of probabilities, at each sample, the probability corresponding to the true class $y_n$ is selected to calculate the individual loss of the sample as given:

$$l_i = -\log(p_{i,y_n}) = -log\left\{\frac{\exp(x_{i,y_n})}{\sum_{j=1}^{C} \exp(x_{i,j})}\right\} \quad (10)$$

For the total $n$ number of samples in a mini-batch, the individual loss is computed for all the samples as a vector $\ell(x, y) = \{l_1, l_2, \dots, l_n\}^{\top}$ from which the individual losses are used in a combination technique (e.g., mean, median, etc.) to calculate the representation of the overall loss, $\mathcal{L}$ of the corresponding mini-batch. In our research, we incorporate a class weighting technique that first computes class inverse weighting (CIW) and then applies a 10-based logarithmic scaling to those weights before integrating the weights into the loss function in Eq. (9). The scaled weights provide an optimized penalty to the minority class if misclassification occurs. CIW is a straightforward way that is calculated by taking the inverse proportionality to the class frequency of that class, calculated by $\frac{\sum n}{f_{y_i}}$, where $f_{y_i}$ represents the class frequency of class $y_i$. However, this weight calculation technique results in high variance among the calculated weights by assigning very low weights for the class with sufficient data, while assigning very large weights for the classes with very few samples. This can lead to reducing the individual performance of a class even though it has sufficient training data. To solve this issue, we applied a logarithmic transformation to this weight calculation method, which is: $w_{y_i} = log(\frac{\sum n}{f_{y_i}})$.

Thus, reducing the range in the scale while keeping the comparative differences. Multiplying this by the original cross-entropy loss of individual samples implies,

$$l'_i = -\log\left(\frac{\sum n}{f_{y_i}}\right).\,log\left\{\frac{\exp(x_{i,y_n})}{\sum_{j=1}^{C} \exp(x_{i,j})}\right\} \quad (11)$$

To calculate the overall loss, we normalize the individual losses in the modified set of losses. Finally, we get our modified loss function as:

$$\mathcal{L}' = -\frac{1}{n}\sum_{1}^{n} \log\left(\frac{\sum n}{f_{y_i}}\right).\,log\left\{\frac{\exp(x_{i,y_n})}{\sum_{j=1}^{C} \exp(x_{i,j})}\right\} \quad (12)$$

### E. EXPLAINABILITY METHODS IN NEUROSLEEPNET

In a transformer network, the encoder part can capture the meaning of each feature sequence ($d_h$) in relation to its context within the temporal sequence ($d_w$). The *MHSA* mechanism in a transformer encoder takes the learned features from the previous layer and uses information from all the features in the sequence to update the feature sequences when required. Thus, the similarity in various convolutional features of different sleep waves becomes distinct, despite their shared characteristics. In our study, we calculated self-attention and visualized the high attention regions to interpret the model, similar to [36]. The primary motivation for extracting self-attention was to identify the regions where the encoder focuses comparatively higher, as self-attention highlights intra-stage contextual dependencies.

Here, each input sequence, $\rho$ has $d_w$ number of token vectors in it, where each vector contains $d_h$ number of values indicating $\rho \in \mathbb{R}^{d_w \times d_h}$. To calculate the self-attentions, we extracted the query ($Q$), key ($K$), and value ($V$) matrices as:

$$Q = \rho W_Q, K = \rho W_k, V = \rho W_V \quad (13)$$

Where $W_Q, W_K, W_V \in \mathbb{R}^{d_h \times d_k}$ are learned projection matrices, and $d_k$ is the dimension of the transformed space. This results in $Q, K, V \in \mathbb{R}^{d_w \times d_k}$. To compute the attention logits between tokens within the same sequence, $\rho$, we used the scaled dot-product attention:

$$\gamma[i,j] = \frac{Q_i \cdot K_j^{+}}{\sqrt{d_k}} \quad (14)$$

Here $\gamma[i,j] \in \mathbb{R}^{d_w \times d_w}$ contains the similarity scores between the *i-th* and *j-th* tokens. These scores are then normalized across the *j-th* dimension with a *softmax* function to get the attention weights:

$$A[i,j] = softmax(\gamma[i,j]) \quad (15)$$

Finally, with these weights, we get the attention output for the *i-th* token,

$$Z_i = \sum_j A[i,j] \cdot V_j \quad (16)$$

Section V-A examines the attention mechanism's influence on identifying sleep stage features and the encoder's contribution to extracting latent representations.

## III. EXPERIMENTS

### A. DATASETS

In our experiment, we used four public datasets on automatic sleep stage scoring. These datasets involve the Sleep-EDF expanded database from PhysioNet, the MESA dataset from NSSR, the Sleep Heart Health Study from NSSR, and the PhysoNet2018 database. The overall summary and splitting ratio of these sleep datasets are shown in Table II. We employed the Sleep-EDFx database as our primary source of various experiments, and also utilized the other databases to obtain the performances.

TABLE II
DATASET DESCRIPTION AND EXPERIMENTAL SETUP

| Name | Subjects | Channel | Scoring | Validation Set | Test Set |
|---|---|---|---|---|---|
| **Sleep-EDFx** | 78 | Fpz-Cz, Pz-Oz | R&K | 0.02% of training data | 21 Subjects (7 each run) |
| **MESA** | 2237 | C4-M1, Oz-Cz, Fz-Cz | AASM | 0.05% of training data | 60 Subjects (20 each run) |
| **SHHS** | 5791 | C3-A2, C4-A1 | R&K | 0.04% of training data | 75 Subjects (25 each run) |
| **Physio 2018** | 994 | F3-M2, F4-M1, C3-M2, C4-M1 | AASM | 0.04% of training data | 60 Subjects (20 each run) |

#### 1) *Sleep-EDF Expanded*

The sleep-EDF expanded (Sleep-EDFx) database is one of the most popular datasets in automatic sleep scoring studies [37-39]. There are 153 whole-night recordings, known as *Sleep Cassette,* that were recorded by 78 healthy Caucasians aged 25, and 101 without any sleep medications. Among the 78 subjects, 75 had recordings from two nights, 2 had recordings only from the first night, and 1 had a recording only from the second night. This resulted in a total of 153 recordings.

#### 2) *Multi-Ethnic Study of Atherosclerosis*

Multi-Ethnic Study of Atherosclerosis (MESA) [40, 41] is set of a clinical-grade PSG recordings that included 2237 sleep participants of diverse ethnic backgrounds, including white, Caucasian, Chinese American, Black, African American, and Hispanic. There were 1198 female participants and 1039 male participants, with ages ranging from 54 to 95. The EEG signals were recorded with a sampling frequency of 256 Hz.

#### 3) *Sleep Heart Health Study*

The sleep recordings of the Sleep Heart Health Study (SHHS) [40, 42] contain PSG recordings from 5791 participants. All participants aged 40 or older have no prior history of sleep apnea treatment. The PSG recordings were sampled at 125 Hz.

#### 4) *PhysioNet2018*

The dataset for the PhysioNet Challenge 2018 was contributed by Massachusetts General Hospital [39, 43]. The dataset consists of 1986 subjects, of whom a total of 994 subjects' recordings were provided to develop and train a sleep scoring system to detect sleep arousal (non-apnea) during sleep. In the training dataset, 67% of the subjects were male, while 33% were female. The average age of the subject was 55, with a standard deviation of 14.3. The signals were recorded with a sampling frequency of 200 Hz.

### B. PREPROCESSING & DATA PREPARATION

We used a 0.5 Hz to 30 Hz band-pass filter to remove signal components from higher frequency bands unrelated to sleep stages. Then, 30-second segments were created, and sleep annotations were included with each epoch. All datasets, except Sleep-EDFx, were resampled to 100 Hz to match the sampling frequency of Sleep-EDFx since NeuroSleepNet was optimized with this dataset. This resampling choice is consistent with prior state-of-the-art studies such as XSleepNet [22] and SleePyCo [19], which also adopted a unified 100 Hz sampling rate across datasets to ensure methodological uniformity and preserve sleep-relevant frequency content (0.5–30 Hz). While creating sleep segments, Z-score normalization was applied to the EEG data.

### C. MODEL SPECIFICATION & CHOICE OF PARAMETERS

In our architecture, the multi-scale convolutional blocks in the *MITCL* module serve primarily as embedding generators. This module extracts local temporal patterns from short windows of EEG data at multiple receptive field sizes while the *MHSA* provides global context modeling. As described in Table I, NeuroSleepNet's *MITCL* is divided into three stages. The smallest scale, i.e., filter size in the convolution block in *MITCL*, was $k_t$=0.25 seconds, which processes $k_t * f_s$ = 0.25*100 = 25 samples at once and slides with a stride of 1. The scale size of the $i^{th}$ block is set as $i * K$ which results in *K*= {25, 50, 100 …}. The number of convolutional blocks (*P*) was determined by a thorough experiment by employing *P* from 1 to 5. The best number of parallel scales was set to *P*=3 by evaluating their performance. Here, the idea of using progressively larger convolutional kernels (e.g., each twice the size of the previous) was previously explored in EEG-based classification models such as [44], and a similar strategy is adopted in our *MITCL* module. A key consideration in this design lies in selecting the *smallest* scale among the multi-scale kernels, as it defines the minimum temporal window available for pattern recognition [45]. This was considered by the temporal characteristics of canonical EEG frequency bands observed during sleep. Sleep EEG primarily contains rhythms in the delta (0.5–3 Hz), theta (4–7 Hz), and alpha (8–13 Hz) ranges. Although convolutional filters are not frequency-selective in the strict sense, they can learn temporal patterns that implicitly correspond to these rhythms — provided their receptive field spans at least one cycle of the underlying oscillation. For example, at a 100 Hz sampling rate, a kernel size of 25 spans 250 ms, which captures one full cycle of a 4 Hz waveform or two cycles of an 8 Hz waveform. Filters shorter than this (e.g., 10 samples = 100 ms) may be insufficient to capture structured rhythmic activity and may instead respond to sharp transients or noise. Models mentioned in [25] with tiny kernel sizes treat the CNN backbone as feature projectors and not temporal analyzers, which might be improper in our transformer-centered model. On the other hand, models such as [10], [20], etc., use large kernels such as fs*4 for longer temporal modeling due to their dependency on larger sequence lengths. In addition, the same filter cannot extract the same information from different datasets if the sampling rate during signal acquisition is different. Thus, during preprocessing, down-sampling was applied to convert those signals that had higher sampling rates to make them similar to our primary experimental dataset, Sleep-EDFx. Therefore, the *T* was 3000 for all the datasets used in our study. To utilize different datasets with different $f_s$ with appropriate *P* and scales *K*, further studies are required. The number of temporal values that were max-pooled was set to $r$ =12 for all the post-convolutional operations.

The *MHSA* module in NeuroSleepNet was configured with two layers (N = 2) for transformer encoders and four attention heads ($n_{head}$ = 4). Using two layers allows the model to capture moderately deep contextual dependencies across the temporal dimension without overfitting, which can occur with deeper stacks, especially when working with short input sequences (30s segments). Four attention heads were chosen to ensure diverse subspace projections while maintaining compatibility with the hidden dimension (2048), as this dimension needs to be evenly divisible by the number of heads.

NeuroSleepNet uses the *ADAM* optimizer with a learning rate of $lr$ = 1e-3 and weight decay regularization with a value of $\gamma$ = 1e-3. For all the datasets, the model was trained with a fixed training batch size of 512 and validated during training with a batch size of 256. The total number of parameters in NeuroSleepNet for our multivariate scheme was $2.12 \times 10^5$ and for the univariate scheme it was $1.11 \times 10^5$.

### D. EXPERIMENTAL DESIGN AND EVALUATION SCHEME

We evaluated our model using a hold-out test set, as specified in Table II for each dataset. The test set was kept completely separate from the training and validation data. We conducted three independent training experiments, each time using a different group of test subjects while training and validating the rest. The final performance is reported as the average over these three test sets. For example, for Sleep-EDFx, we tested on 21 unique subjects in total (7 per run). The approach was applied to the other datasets as well, with test set sizes shown in Table II. The evaluation metrics we used in our study include accuracy, macro-F1 score, and Cohen's kappa score. We also showed the confusion matrix for five-class classification to represent how well our model performs for each task with per-class recall scores. To measure the fraction of correctly classified sleep stages out of all five classes, we calculated accuracy. The balance accuracy was calculated to account for class imbalance issues by averaging the recall score for each stage of sleep.

TABLE III
DATASET DESCRIPTION AND EXPERIMENTAL SETUP

| Model | Overall Performance | | | Per-Class F1-Score | | | | |
|---|---|---|---|---|---|---|---|---|
| | Acc. | mF1 | κ | W | N1 | N2 | N3 | R |
| Sleep-EDFx | 86.1 | 80.8 | 0.805 | 91.7 | 48.5 | 89.3 | 88.7 | 84.6 |
| MESA | 82.0 | 76.3 | 0.753 | 91.6 | 48.8 | 85.3 | 80.0 | 76.8 |
| Physio2018 | 80.5 | 76.8 | 0.768 | 81.9 | 58.1 | 83.0 | 81.5 | 79.5 |
| SHHS | 86.7 | 80.9 | 0.80 | 85.5 | 52.5 | 90.0 | 89.9 | 86.6 |

## IV. EXPERIMENTAL RESULTS

This section presents the results obtained from a series of experiments. In Section IV-A, we report the overall performance of NeuroSleepNet across all four datasets. The majority of the architectural design was independent of dataset-specific optimization, as core design choices such as module arrangement, convolutional kernel design principles, and feature representation strategies were pre-established based on the design scheme outlined in Sections II-B, II-C, and II-D. It is important to note that the tunable parameters of the model were optimized solely using the Sleep-EDFx dataset, while the remaining datasets were used for independent evaluation to assess generalizability.

### A. OVERALL PERFORMANCE OF NEUROSLEEPNET

The performance of the NeuroSleepNet model was validated using four publicly available datasets, as previously mentioned, following our proposed one-to-one scheme. In this approach, the overall accuracy, MF1 score, and Cohen's kappa value were 86.1 ± 0.6%, 80.8 ± 0.5%, and 0.805 ± 0.008 for the Sleep-EDF extended dataset; 82.0 ± 0.8%, 76.3 ± 0.7%, and 0.753 ± 0.010 for MESA; 80.5 ± 0.4%, 76.8 ± 0.5%, and 0.738 ± 0.006 for Physio2018; and 86.7 ± 0.3%, 80.9 ± 0.4%, and 0.804 ± 0.004 for SHHS, respectively. The overall performance is summarized in Table III, while the confusion matrix for NeuroSleepNet is presented in Fig. 2. These results demonstrate that, without dataset-specific parameter optimization, the model exhibits robust performance across datasets of varying sizes and characteristics. Furthermore, a hypnogram plot comparing the model's predictions with annotations scored by a sleep expert is presented in Fig. 3, visually demonstrating NeuroSleepNet's ability to closely replicate human expertise in sleep staging with high precision throughout the night.

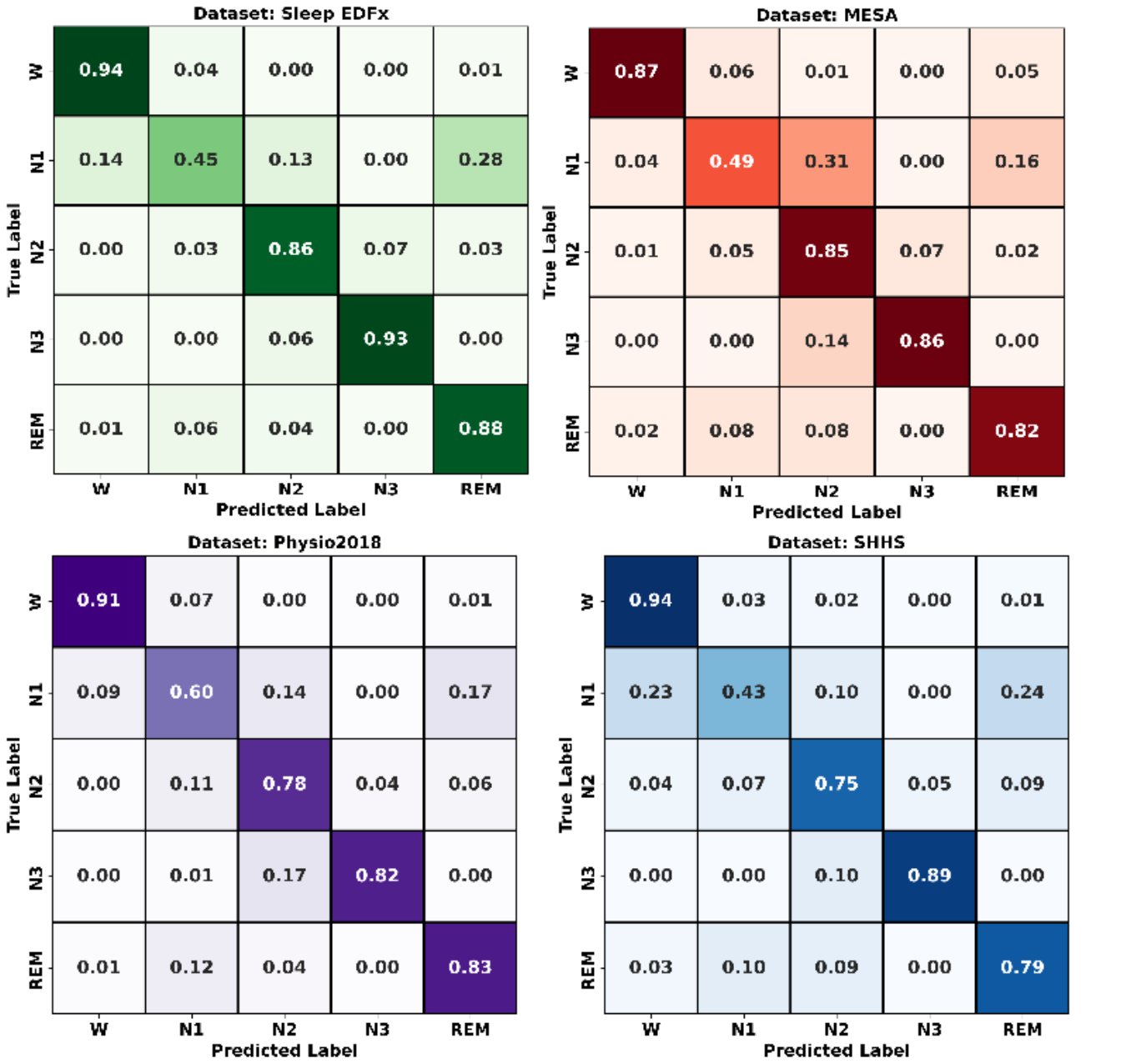

FIGURE 2. The confusion matrices of NeuroSleepNet on four datasets: Sleep-EDFx, MESA, Physio2018, and SHHS. The actual class is positioned vertically, and the predicted class is positioned horizontally.

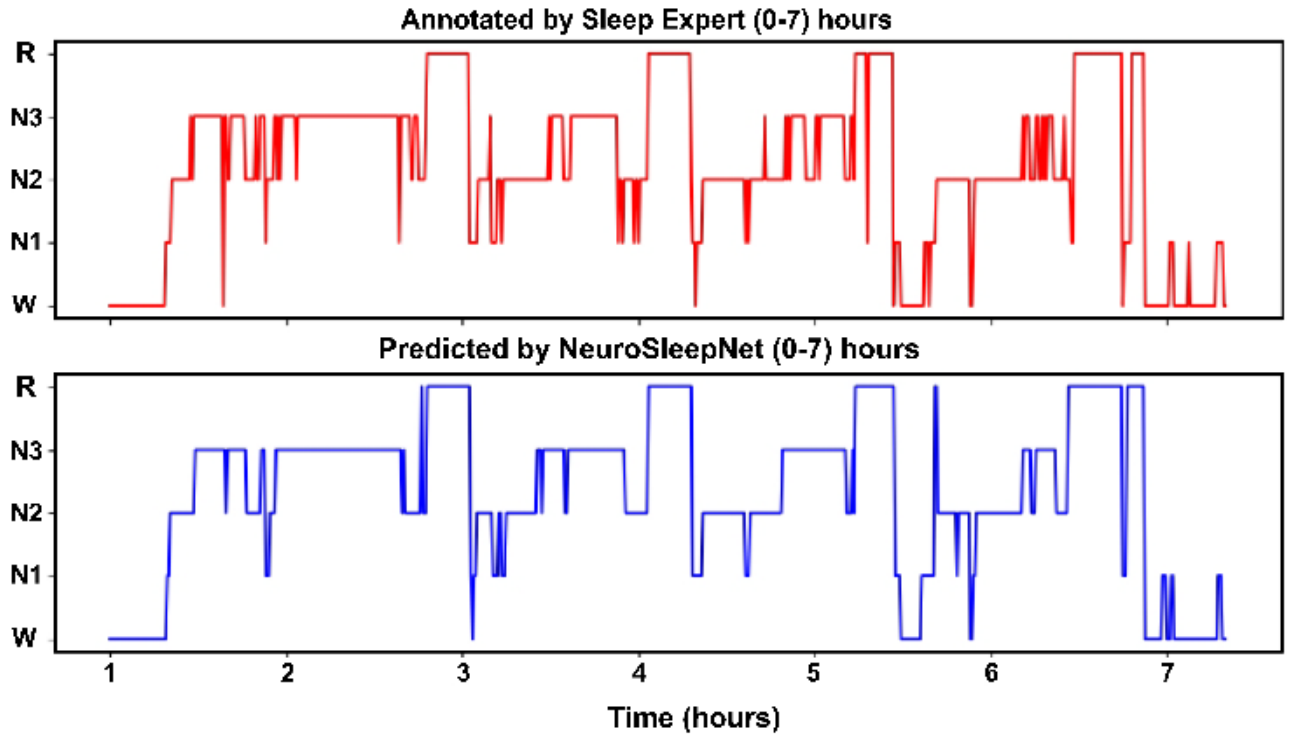

FIGURE 3. Two hypnograms comparing the whole night sleep stage scoring done by a sleep expert (top) and NeuroSleepNet (bottom). The comparison is shown from the recordings of subject 01 from the Sleep-EDFx dataset. For this subject, the scores are acc: 86.4, Mf1: 81.3, and $\kappa$: 81.9.

## V. DISCUSSIONS

### A. TRANSFORMING CONVOLUTIONAL FEATURES INTO CONTEXUAL FEATURES: IMPACT ON MODEL DECISION-MAKING

In this section, we explain the attention mechanism in NeuroSleepNet concerning sleep stage characteristics and how the encoder contributes to capturing subtle, hidden information. The *MHSA* module in NeuroSleepNet offers a direct pathway for interpreting the model's decision-making by highlighting high-attention regions from the static input features. Sleep signals often contain physiological markers (e.g., sleep spindles, K-complexes) that human experts rely on for stage classification. While the goal of this analysis was not to explicitly detect these markers, the key motivation was to examine whether the *MHSA* module naturally attends to similar regions that sleep experts consider important during manual scoring. This alignment provides an interpretable basis for the attention patterns generated by the *MHSA* encoders in NeuroSleepNet. Here, we have shown two examples for three stages, notably N2, REM, and N1. Since wake and N3 are characterized by apparent spectral and amplitude characteristics and are easily identifiable via simple methods such as short-term Fourier transform (STFT), we did not show them here. The micro-events identified in the signals were analyzed using three sources: expert-assigned labels, AASM scoring guidelines, and STFT analysis. These were then compared against attention heatmaps generated by the *MHSA* module to evaluate whether the model focuses on regions relevant to stage classification. In Fig. 4(a), the input signal (left) is labeled as stage N2 by a sleep expert, with a sleep spindle clearly visible (highlighted in red). Sleep spindles are defined by AASM as bursts of 11–16 Hz sinusoidal activity lasting 0.5 to 1.5 seconds. The corresponding *MHSA* heatmap (right) shows a clear concentration of attention in this spindle region, indicating that the model is sensitive to this important micro-event.

Fig. 4(b) shows another N2 segment containing a K-complex - a sharp negative wave followed by a slower positive component lasting at least 0.5 seconds. Similar to Fig. 4(a), the *MHSA* module assigns high attention to this region. This suggests that it recognizes the K-complex as a salient feature. In Figs. 4(c) and 4(d), we present REM-labeled signals, where rapid eye movements (REMs) are visible, characterized by sharply peaked deflections under 0.5 seconds, in line with AASM criteria. The *MHSA* module exhibits localized attention in these regions as well, indicating alignment with domain-relevant cues. Two examples of the N1 stage are also shown. In Fig. 4(e), a correctly classified 30-second N1 segment exhibits widely distributed attention across the signal. This is expected, as N1 is often difficult to classify and may require the model to attend to mixed-frequency patterns or slow eye movements. In contrast, Fig. 4(f) shows an N1 segment misclassified as N2. The *MHSA* heatmap here reveals high attention on a region that may resemble a K-complex to the model, though this is not a true K-complex, leading to incorrect classification. These examples show that NeuroSleepNet's *MHSA* module often focuses on important physiological features like spindles, K-complexes, and eye movements, which are also used by human experts to score sleep stages. This suggests that the model learns meaningful patterns from the data. However, some misclassifications reveal that the model can also focus on misleading signal patterns, especially in harder-to-classify stages like N1. Overall, attention heatmaps help us understand when the model is making correct decisions and where it might go wrong.

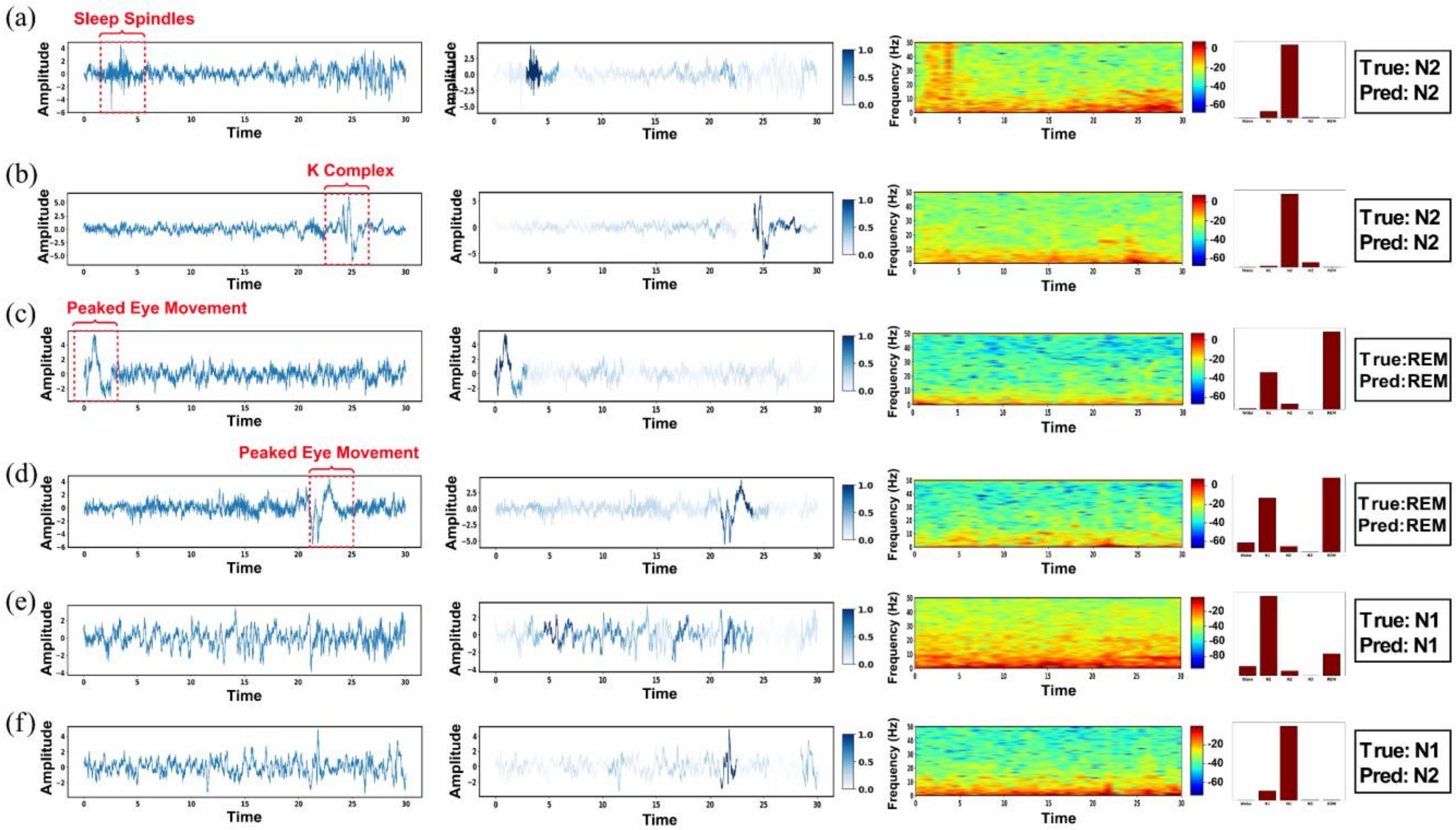

**FIGURE 4.** Examples of N2, REM, and N1 stages showing physiological markers (per AASM guidelines and STFT analysis) alongside high-attention regions from NeuroSleepNet. The highlighted areas align with characteristic features used in manual scoring, illustrating the interpretability of the MHSA module. Signals are from Subject 1 (Sleep-EDFx, excluded from training/validation). Softmax outputs are also shown for reference.

### B. WEIGHTED COST FUNCTION COMPARED WITH OTHER CONVENTIONAL APPROACHES

The impact of the class imbalance problem is severe and reduces the learning capability of the minority classes. One of the major methods for dealing with class imbalance problems is data augmentation, which expands datasets by transforming existing samples. A key challenge with data augmentation is that it requires additional computation to generate new samples. In this study, we focused on a weighting strategy to address class imbalance. There can be two primary approaches: dynamic weighting, such as focal loss, and static weighting, such as balanced weights. We chose static weighting over dynamic weighting due to its simplicity, fewer hyperparameters, and more stable training behavior. Balanced weight (BW) has also been previously applied in sleep stage scoring, as seen in [12] and [46]. Both the BW and our proposed log-scale-based weights are derived from the raw CIW scheme. The traditional CIW measures the number of samples in a class and calculates a weight inversely proportional to that number. However, the deviation from the minimum weight to the maximum is very high. The CIW calculation can potentially reduce the learning capability of samples with sufficient training data, as shown in Fig. 5. The figure shows that for CIW, performance increases for classes with lower sample sizes, such as N1 and N3. However, the performance decreases with classes with a high number of training samples, such as W, N2, and REM. The static BWs, on the other hand, are calculated as follows:

$$w_m = \frac{\sum n}{f_{y_i}} \times \frac{1}{C} \quad (17)$$

where $w_m$ denotes the weight of the $m$-th class, $N$ indicates the number of samples in the dataset, $n_i$ is the number of samples in the $m^{th}$ class, and $Y$ is the number of classes [47]. This method produces weights with lower variance and deviation than the CIW. However, this method can also result in disproportionate penalization of the classes with sufficient data. To better illustrate the difference between balanced weights and our proposed logarithmic scaling method, consider a batch of 50 sleep epochs with class distribution: {W: 10, N1: 5, N2: 20, N3: 10, REM: 5}. According to the balanced weight formulation, the resulting weights are approximately $\{1.0, 2.0, 0.5, 1.0, 2.0\}$. In contrast, applying logarithmic scaling on the class-inverse frequencies yields weights of approximately $\{1.6, 2.3, 0.9, 1.6, 2.3\}$. In this case, both methods increase the weight of the rare classes (N1, REM) while decreasing the weight of the majority class (N2). However, logarithmic scaling achieves this more conservatively: it avoids excessively diminishing the loss contribution from the majority class while still amplifying the penalty for minority class errors. This approach enables continued learning from dominant classes, without overwhelming the model's focus.

In a more extreme case, e.g., {W: 10, N1: 5, N2: 200, N3: 10, REM: 5}, balanced weights yield {4.6, 9.2, 0.23, 4.6, 9.2}, whereas log-scale weights give {3.1, 3.8, 0.14, 3.1, 3.8}. Though both methods decrease the weight of the overrepresented N2 class, log-scaling preserves a smaller variance between underrepresented classes (W, N1, N3, REM), reflecting the fact that they are all similarly scarce. This highlights a key benefit of the log-scaled strategy: in highly skewed settings, it reduces the *relative variance* between minority classes, treating them with more uniform importance while still reducing the majority class's influence.

In Fig. 5, it can be observed that while BW enhances N1 performance, it significantly decreases the per-class performance of W, N, and N3. The proposed log-scale weight assignment technique applies a logarithmic transformation to CIW. Logarithmic transformations are widely utilized in statistical analysis to address skewed distributions. Since CIW assigns very high weights to the minority class and very low weights to the majority class, the log transformation ensures that minority classes receive appropriate importance without dominating the training process. As shown in Fig. 5, our log scale-based weight assignment method outperforms other techniques, including training with no weights. While the improvement in REM stages is small and a negligible amount of decrease in N2, our weight assignment technique significantly enhances the W, N1, and N3 classes. If no weights are assigned, the loss function by default considers all equal weights of 1, resulting in a variance of 0. This does not prioritize the minority class, such as N1 or N3, during training. Since the CIW results in very high variance as it assigns very large weights to classes with low samples, we can observe some improvement in minority classes such as N1 in Fig. 5. The balanced weights, in contrast, typically have lower variance than CIW. In the future, we aim to combine the proposed log-scaled weighting scheme with focal loss to evaluate its effectiveness under dynamic weighting strategies.

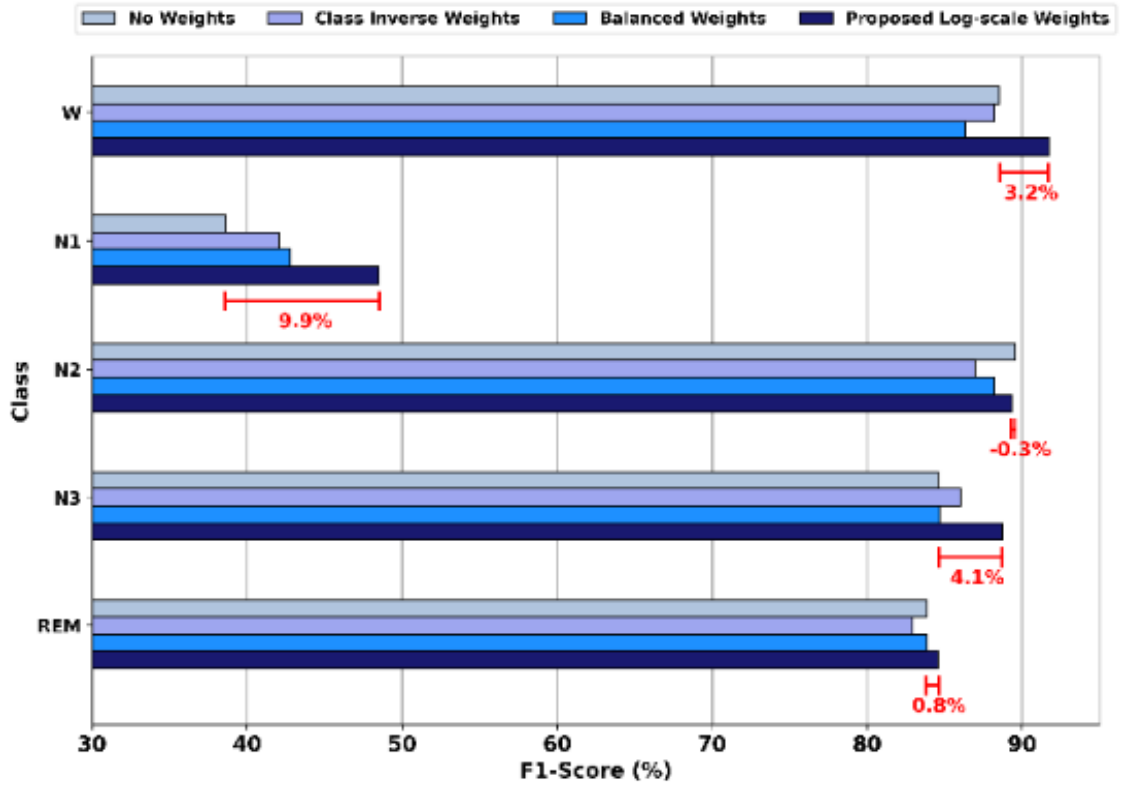

FIGURE 5. The per-class F1 scores for five different classes are compared across three weight assignment techniques for the loss function. The log-scale-based weight assignment method demonstrated higher per-class F1 scores for almost all classes, indicating a more balanced and improved classification performance. The difference in F1 scores between the absence of weighting and the proposed log-scale weighting is also highlighted using difference markers. The most notable improvement is observed in the N1 class, with a 9.9% increase in the F1-score, while the N2 class shows a minimal decrease of only 0.3%.

### C. PERFORMANCE COMPARISON WITH PREVIOUS STATE-OF-THE-ART MODELS AND FUTURE WORKS

The performance of NeuroSleepNet is shown in Table IV and compared with the existing SOTA deep learning models for automated sleep scoring. In the comparison, we included the models' performances, input type, and sequence length simultaneously. The comparison shows that our model achieved state-of-the-art performance with only a single sequence length, which is a significant advantage compared to the existing models that used multiple sequences. Additionally, the longer sequences increase computational demands, which is also minimized in our case. Furthermore, the proposed method does not require any hand-crafted features or time-frequency image representation of the raw signal. The existing baseline SOTA study is based on a multi-view scheme that uses both raw signal and time-frequency images as their input [22, 30]. This time-frequency conversion from the raw signal also increases the computational burden. In contrast, our proposed method is designed to achieve comparable SOTA performance with only the raw signal. The other methods that achieved comparable performance with raw signals either require a higher number of input epochs compared to ours [19] or require both previous and subsequent sleep segments [33]. As reported earlier, NeuroSleepNet uses $2.12 \times 10^5$ parameters. On the other hand, the two high-performing SOTA models such as, [22] uses $5.6 \times 10^6$ parameters, which is around 26 times higher than ours, and [19] uses $2.01 \times 10^6$ parameters, which is around 10 times higher than our model. Also, in SleePyCo, the convolutional backbone utilizes around 60% of the whole network, contributing to higher computational complexity. On the other hand, the *CSF, MITCL*, and *PCC* together contribute to only 3% of the overall model parameters, which indicates NeuroSleepNet's representation learning as a more streamlined approach for sleep stage classification. Additionally, we have also evaluated the computational efficiency of NeuroSleepNet using Floating Point Operations (FLOPs), which measure the number of mathematical operations performed during inference or training - directly reflecting a model's computational cost. Generally, models with fewer FLOPs require less computational power and run faster, making them more suitable for real-time applications. NeuroSleepNet demonstrates significant computational efficiency, requiring only $1.33 \times 10^7$ FLOPs, whereas [22] and [19] require $1.58 \times 10^8$ and $1.45 \times 10^{10}$ FLOPs, respectively, while maintaining competitive classification performance. In addition, NeuroSleepNet's transformer encoder module is designed to operate with a lower input embedding dimension, which significantly reduces the parameter count of the overall model [48]. For example, in [16], embeddings are extracted from time-frequency representations of EEG signals with an embedding dimension $d_{model}$ of 128, leading to a parameter count of the transformer module of approximately $2.6 \times 10^6$ from a total parameter count of 3.7e6. On the other hand, models such as [19] use default parameters as depicted in [35] as a $d_{model}$ of 128 that resulted in $5.8 \times 10^5$ parameters. In contrast, NeuroSleepNet's efficient representation learning yields a compact embedding dimension of $d_{model}$=16 and employs fewer heads and encoder layers, resulting in only approximately $1.3 \times 10^5$ parameters in the Transformer encoder. This design choice demonstrates that NeuroSleepNet can serve as a more efficient Transformer-based architecture, achieving a lower parameter count in the transformer encoder module while maintaining strong performance on 30-second EEG segments.

Thus, we argue that as a one-to-one approach, NeuroSleepNet's compact design and explainability make it a more practical choice for sleep scoring. Phan et al. [29] and Fiorillo et al. [49] both highlight the need for lightweight, interpretable, and clinically usable sleep staging models. NeuroSleepNet directly aligns with these goals through its single-epoch, low-latency architecture and attention-based interpretability. While prior work emphasizes leveraging long temporal context, our approach explores how reducing context can still preserve performance without sacrificing relevant information. We hypothesize that one reason many sleep staging models struggle to generalize across diverse cohorts is their heavy reliance on long-range temporal context. Sleep patterns—both across full cycles and within partial sequences—can vary significantly from person to person, often leading to irregular or disrupted stage transitions. In such cases, models that rely heavily on consistent long-term patterns may underperform. Thus, even well-designed sleep stage classifiers can face generalization challenges if they do not explicitly account for variability in sequence structure across individuals and populations [50]. Since NeuroSleepNet classifies each sleep stage based solely on a single epoch, variability in sleep patterns or disruptions in stage transitions and multi-channel settings, as demonstrated in section IV-B, makes it more flexible. However, it is important to note that while datasets like SHHS and MESA provide clinical-grade PSG recordings and offer a strong foundation for model benchmarking, we acknowledge that real-world clinical deployment presents additional challenges. Variations in recording devices, acquisition protocols, scorer interpretations, and patient comorbidities could affect model generalizability. Therefore, future validation on prospectively collected clinical datasets remains an important step to fully establish the robustness of NeuroSleepNet in diverse clinical environments. While NeuroSleepNet is not specifically designed for pathological sleep stage classification, its modular design allows for flexible adaptation. This would enable evaluation on cohorts with diverse pathologies, such as narcolepsy, to assess generalizability.

TABLE IV
PERFORMANCE COMPARISON BETWEEN NEUROSLEEPNET (MULTIVARIATE) AND OTHER STATE-OF-THE-ART DEEP LEARNING BASED METHODS FOR AUTOMATIC SLEEP SCORING

| System | | | Overall Performance | | | | Per-class F1 Score | | | | |
|---|---|---|---|---|---|---|---|---|---|---|---|
| **Database** | **Model** | **Training Param.** | **Current (Previous)** | **Acc.** | **MF1** | **κ** | **W** | **N1** | **N2** | **N3** | **REM** |
| **Sleep-EDFx** | **NeuroSleepNet** | **~ 0.2M** | **1(0)** | **86.1** | **80.8** | **0.805** | 91.7 | 48.5 | **89.3** | **88.7** | **84.6** |
| | SleePyCo [19] | ~ 2.0M | 1(9) | 84.6 | 79.0 | 0.787 | **93.5** | 50.4 | 86.5 | 80.5 | 84.2 |
| | XSleepNet [22] | ~ 5.6M | 1(19) | 84.0 | 77.9 | 0.778 | -- | -- | -- | -- | -- |
| | Korkalainen et al. [14] | ~ 1.7M | 1(99) | 83.7 | -- | 0.770 | -- | -- | -- | -- | -- |
| | TinySleepNet [13] | ~ 1.3M | 1(14) | 83.1 | 78.1 | 0.770 | 92.8 | **51.0** | 85.3 | 81.1 | 80.3 |
| | SeqSleepNet [9] | ~ 0.2M | 1(29) | 82.6 | 76.4 | 0.760 | -- | -- | -- | -- | -- |
| | SleepTransformer [16] | ~ 3.7M | 1(20) | 81.4 | 74.3 | 0.743 | 91.7 | 40.4 | 84.3 | 77.9 | 77.2 |
| | U-Time [26] | ~ 1.2M | 1(34) | 81.3 | 76.3 | 0.745 | 92.0 | 51.0 | 0.84 | 0.75 | 0.80 |
| | AttnSleep [20] | ~ 0.1 M | 1(79) | 81.3 | 75.1 | 0.740 | 92.0 | 42.0 | 85.0 | 82.1 | 74.2 |
| | DeepSleepNet-Lite | ~ 0.6M | 1(2) | 80.9 | 76.7 | 0.740 | 90.7 | 48.0 | 84.4 | 79.8 | 79.7 |
| | SleepEEGNet [12] | ~ 2.6M | -- | 80.0 | 73.5 | 0.730 | 91.7 | 44.0 | 82.5 | 73.5 | 76.1 |
| **Physio2018** | **NeuroSleepNet** | **~ 0.2M** | **1(0)** | 80.5 | 76.8 | **0.738** | 81.9 | 58.1 | 83.0 | **81.5** | 79.5 |
| | SleePyCo [19] | ~ 2.0M | 1(9) | **80.9** | **78.9** | 0.737 | **84.2** | **59.3** | **85.3** | 79.4 | **86.3** |
| | XSleepNet [22] | ~ 5.6M | 1(19) | 80.3 | 78.6 | 0.732 | -- | -- | -- | -- | -- |
| | SeqSleepNet [9] | ~ 0.2M | 1(29) | 79.4 | 77.6 | 0.719 | -- | -- | -- | -- | -- |
| | U-Time [26] | ~ 1.2M | 1(34) | 78.8 | 77.4 | 0.714 | 82.5 | 59.0 | 83.1 | 79.0 | 83.5 |
| **SHHS** | **NeuroSleepNet** | **~ 0.2M** | **1(0)** | 86.7 | 80.9 | 0.804 | 85.5 | **52.5** | **90.0** | **89.9** | 86.6 |
| | L-SeqSleepNet [30] | ~ 0.6M | 1(200) | 88.4 | **81.4** | **0.830** | **93.1** | 51.1 | 89.0 | 84.9 | **89.8** |
| | SleePyCo [19] | ~ 2.0M | 1(9) | **87.9** | 80.7 | **0.830** | 92.6 | 49.2 | 88.5 | 84.5 | 88.6 |
| | SleepTransformer [16] | ~ 3.7M | 1(20) | 87.7 | 80.1 | 0.828 | 92.2 | 46.1 | 88.3 | 85.2 | 88.6 |
| | XSleepNet [22] | ~ 5.6M | 1(19) | 87.6 | 80.7 | 0.826 | 92.0 | 49.9 | 88.3 | 85.0 | 88.2 |
| | Sors et al. [8] | ~ 3.3M | 1(2), 1 future | 86.8 | 78.5 | 0.850 | 91.4 | 42.7 | 88.0 | 84.9 | 85.4 |
| | IITNET [25] | ~ | 1(9) | 86.7 | 79.8 | 0.812 | 90.1 | 48.1 | 88.4 | 85.2 | 87.2 |
| | SeqSleepNet [9] | ~ 0.2M | 1(19) | 86.5 | 78.5 | 0.810 | -- | -- | -- | -- | -- |

### D. ABLATION STUDIES

To assess the contribution of each component in NeuroSleepNet, we conduct an ablation study by systematically removing or modifying key architectural elements. We evaluate the model performance on the Sleep-EDFx dataset using accuracy, macro-F1 score, and Cohen's kappa as primary metrics. The ablations include:

1. **NeuroSleepNet-NSF (No Spatial Filtering):** To verify the effectiveness of cross-channel learning via convolutional spatial filtering, we removed the spatial filtering layer and fed the raw EEG signal directly into the multi-scale temporal context learning (*MITCL*) module.
2. **NeuroSleepNet-SC (Single-scale CNN):** To verify the effectiveness of the *MITCL* module, we replaced it with a single-scale convolutional layer and restricted the network to learn temporal features at only one fixed resolution.
3. **NeuroSleepNet-NTE (No Transformer Encoder):** To evaluate the impact of the transformer-based self-attention mechanism, we removed the transformer encoder and directly fed the feature representations from *MITCL* into the classification layer.
4. **NeuroSleepNet-NWL (No Weighted Loss):** To determine the effect of the logarithmic-scaled weighted loss function on class balancing, we replaced it with a standard cross-entropy loss.

**5. NeuroSleepNet-MTCL:** To compare the impact of regular multi-scale temporal context learning (MTCL) against our proposed multi-scale independent temporal context learning (MITCL), we constructed a variant of the model where the MITCL module was replaced with standard 1D convolutional layers. In this variant, the kernel was not constrained along the spatial (channel) dimension, allowing the filters to aggregate information across channels. All other hyperparameters, including the number of filters, kernel sizes, and architectural settings, were kept identical to the original MITCL configuration to ensure a fair comparison. The obtained performance from these variants is outlined in Table V. The performance highlights the importance of each component in NeuroSleepNet. The largest performance drop occurs when removing the transformer encoder (NeuroSleepNet-NTE), leading to a decrease in accuracy of 7.7%, macro-F1 score of 9.8%, and kappa value of 0.114. This variant also exhibits a drastic decline in per-class F1 scores for N1 and REM which indicates the crucial role of *MHSA* in capturing contextual dependencies in NeuroSleepNet. The removal of multi-scale temporal context learning (NeuroSleepNet-SC) also results in a notable decline in performance. While the absence of convolutional spatial filtering (NeuroSleepNet-NSF) slightly reduces accuracy and balanced classification, it still outperforms the no-transformer model. Additionally, removing the weighted loss function (NeuroSleepNet-NWL) with standard categorical loss function negatively impacts classification balance, particularly for underrepresented classes like N1 and REM. These findings validate our architectural design choices and emphasize the necessity of each module in achieving state-of-the-art performance in automatic sleep staging.

TABLE VI

VALIDATION OF DIFFERENT VARIANTS OF NEUROSLEEPNET

| Model | Overall Performance | | | Per-Class F1-Score | | | | |
|---|---|---|---|---|---|---|---|---|
| | Acc | F1 | κ | W | N1 | N2 | N3 | R |
| NeuroSleep Net-NSF | 82.1 | 76.7 | 0.75 | 87.5 | 42.0 | 88.6 | 88.2 | 80.7 |
| NeuroSleep Net-SC | 80.0 | 75.3 | 0.75 | 88.7 | 37.4 | 87.6 | 86.2 | 80.6 |
| NeuroSleep Net-NTE | 78.4 | 71.0 | 0.69 | 86.1 | 27.4 | 83.4 | 82.8 | 75.2 |
| NeuroSleep Net-NWL | 82.3 | 74.3 | 0.73 | 87.6 | 33.7 | 86.4 | 80.5 | 83.5 |
| NeuroSleep Net-MTCL | 80.0 8 | 75.5 | 0.72 | 89.8 | 41.2 | 82.7 | 84.1 | 79.8 |
| **NeuroSlee pNet** | **86.1** | **80.8** | **0.80** | **91.7** | **48.5** | **89.3** | **88.7** | **84.6** |

## VI. CONCLUSION

We introduced NeuroSleepNet, a novel automatic sleep scoring architecture based on EEG that utilizes spatial and multi-scale independent temporal context learning feature extractor and multi-head self-attention transformer encoders to classify sleep stages with minimal sequence length. We also modified the loss function with a log-scaled weight assignment technique to achieve a more balanced performance. NeuroSleepNet provides comparable performance with the current SOTA architecture by using only a single input epoch and excluding the requirement of transition rules. The model accommodates both univariate and multivariate inputs. Furthermore, we demonstrated how NeuroSleepNet can serve as an explainable AI model, offering insights into its decision-making process for sleep stage classification. NeuroSleepNet has significant potential for real-time automatic sleep scoring. The explainability techniques introduced in this research can reveal how deep learning models focus on specific sleep signal regions. This can be used to compare model attention with human expert annotations and assess whether the model prioritizes regions similar to those of sleep experts. It also helps to identify subtle EEG patterns that contribute to a deeper understanding of sleep-related characteristics. As a natural extension of this work, we plan to apply NeuroSleepNet's principles for physiological signals extracted from wearables to monitor sleep stages and sleep-related disorders.

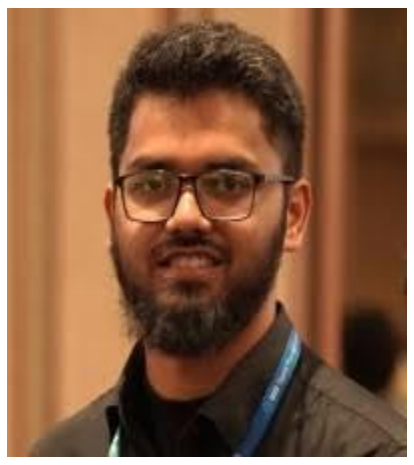

**MUHAMMAD SUDIPTO SIAM DIP** received the B.Sc. degree in Electrical and Electronic Engineering from the Faculty of Electrical and Computer Engineering, Rajshahi University of Engineering and Technology (RUET), Rajshahi, Bangladesh. His current research interests include machine learning, deep learning based biomedical and time-series signal processing. He has become the 3rd prize winner of IEEE Signal Processing Cup 2024, ICASSP.

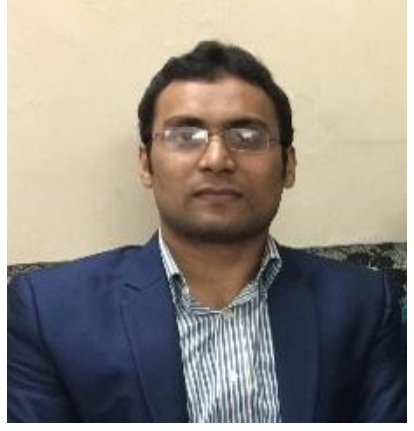

**MOHAMMOD ABDUL MOTIN** (Senior Member, IEEE) is currently working as an Associate Professor in the Department of Electrical & Electronic Engineering, Rajshahi University of Engineering & Technology, Bangladesh. He completed his Ph.D. degree with the Department of Electrical and Electronic Engineering at the University of Melbourne, Australia, in 2020. Prior to that, he received his M.Sc. Eng. and B.Sc. Eng. degrees in electrical & electronic engineering from the Rajshahi University of Engineering & Technology, Rajshahi, Bangladesh, in 2014 and 2011, respectively. His research interests include biomedical signal processing and modeling, time series analysis, machine learning, embedded systems, monitoring neurological disorders, and sensor devices for wearable health care monitoring. Dr. Motin is the recipient of several awards and prizes, including the IEEE Upsilon Pi Epsilon Award, IEEE Lance Stafford Larson Award, IEEE Richard E. Merwin Scholarship, W. E. and C. H. H. Cook Memorial Award, ISSNIP Travel Grant, Melbourne Abroad Travelling Scholarship (MATS), ANZSCON 2017 Travel Grant, IEEE R10 SAC Best Paper Award (PG Level), IEEE Australia Council Student Paper Award (Post-graduate), Best Paper Award in ICEEE 2021, Best Paper Award in ICEEICT 2024, and the University of Melbourne MIRS and MIFRS Scholarships.

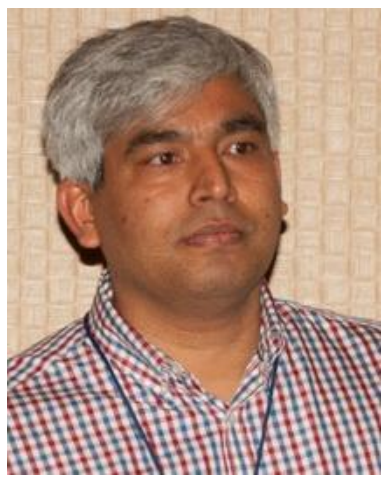

**CHANDAN KARMAKAR** (Member, IEEE) received the Ph.D. degree from The University of Melbourne, Melbourne, Australia. He is an Associate Professor with the School of Information Technology, Deakin University, Melbourne. He has published one book and more than 170 research articles, including 90 plus journal articles. His research interests include applied artificial intelligence, internet of medical things, wearable devices, signal quality analysis, time series analysis, nonlinear signal processing, irregularity and complexity, health informatics and modeling of physiological systems.

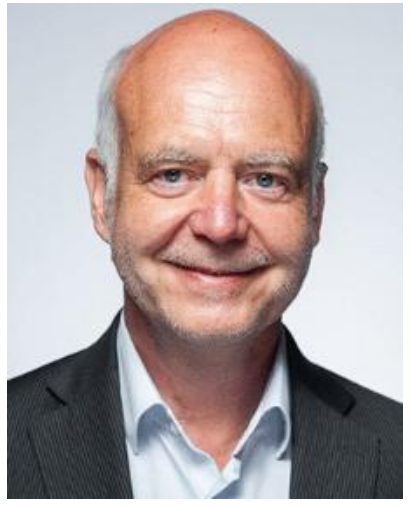

**THOMAS PENZEL** (Fellow, IEEE) graduated in physics, human biology, and physiology from the University Marburg, Marburg, Germany. In 2006, he moved to Berlin as the Director of research of the Interdisciplinary Sleep Medicine Center, Charité - Universitätsmedizin Berlin, Berlin, Germany. He is currently the President of the German Sleep Society, President-Elect of World Sleep Society (WSS), and IEEE EMB Vice-President publications. He authored many publications with a focus on sleep apnea and biomedical engineering. He is the Editor-in-Chief of the journal Sleep and Breathing, and editorial board member of other journals.

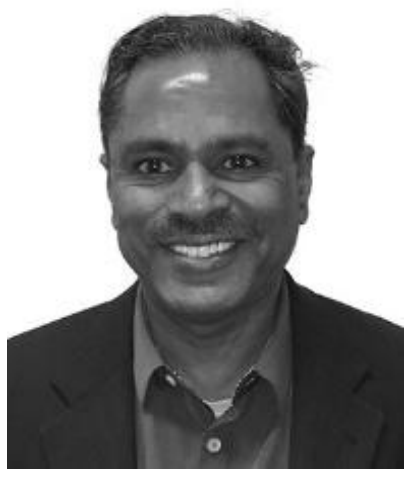

**MARIMUTHU PALANISWAMI** (Life Fellow, IEEE) received the M.E. degree from the Indian Institute of Science Bengaluru, Bengaluru, India, in 1977, the M.Eng.Sc. degree from The Uni- versity of Melbourne, Parkville, Melbourne, VIC, Australia, in 1983, and the Ph.D. degree from the University of Newcastle, Callaghan, NSW, Australia, in 1987. He is currently a Professor with the Department of Electrical and Electronic Engineering, The University of Melbourne. He has published over 400 refereed research articles and led one of the largest funded Australian research councils, a research network on intelligent sensors, sensor networks, and information processing programs. He has been funded by several ARC and industry grants (more than $60M) to conduct research in support vector machines, sensors, sensor networks, the IoT, machine learning, neural networks, pattern recognition, signal processing, and control areas. He represents Australia as a core partner in European Union FP7 projects, such as SENSEI, Smart Santander, the Internet of Things Initiative, and SocIoTal.